\documentclass[12pt,a4paper]{article}

\usepackage[english]{babel}

\usepackage{tipa}
\usepackage{url}

\usepackage{caption}

\usepackage{amsmath}
\usepackage{amsfonts}
\usepackage{amssymb}

\usepackage{natbib}

\usepackage[utf8]{inputenc}

\usepackage{graphicx}

\def\bit  {\begin{itemize}}
\def\eit  {\end{itemize}}
\def\beq  {\begin{equation}}
\def\eeq  {\end{equation}}
\def\beqa {\begin{eqnarray}}
\def\eeqa {\end{eqnarray}}
\def\bef  {\begin{figure}}
\def\ef   {\end{figure}}

\def\ben  {\begin{enumerate}}
\def\een  {\end{enumerate}}
\def\bec  {\begin{center}}
\def\ec   {\end{center}}
\def\bet  {\begin{tabbing}}
\def\et   {\end{tabbing}}
\def\vs   {\vspace{0.5cm}}

\title{\vs\vs\vs\vs\vs\vs Perception of categories: from coding efficiency to reaction times \vs}
\author{Laurent Bonnasse-Gahot$^{1}$ and Jean-Pierre Nadal$^{1,2,\ast}$}

\date{\vs
{(1) Centre d'Analyse et de Math\'ematique Sociales \\(CAMS, UMR 8557 CNRS -- EHESS)} \\
{\'Ecole des Hautes \'Etudes en Sciences Sociales, Paris, France} \\
{(2) Laboratoire de Physique Statistique \\(LPS, UMR 8550 CNRS -- ENS -- UPMC -- Paris Diderot)} \\
{\'Ecole Normale Sup\'erieure, Paris, France}\\
{\vs\vs\vs $\ast$ corresponding author - email: nadal@lps.ens.fr}}

\begin{document}

\maketitle

\newpage
\begin{center}
\textbf{Abstract}
\end{center}
Reaction-times in perceptual tasks are the subject of many experimental and theoretical studies.
With the neural decision making process as main focus, most of these works concern discrete (typically binary) choice tasks, implying the identification of the stimulus as an exemplar of a category.
Here we address issues specific to the perception of categories (e.g. vowels, familiar faces, ...), making a clear distinction between identifying a category (an element of a discrete set) and estimating a continuous parameter (such as a direction). 
We exhibit a link between optimal Bayesian decoding and coding efficiency, the latter being measured by the mutual information between the discrete category set and the neural activity. We characterize the properties of the best estimator of the likelihood of the category, when this estimator takes its inputs from a large population of stimulus-specific coding cells. 
Adopting the diffusion-to-bound approach to model the decisional process, this allows to relate analytically the bias and variance of the diffusion process underlying decision making to macroscopic quantities that are behaviorally measurable.
A major consequence is the existence of a quantitative link between reaction times and discrimination accuracy.
The resulting analytical expression of mean reaction times during an identification task
accounts for empirical facts, both qualitatively (e.g. more time is needed to identify a category from a stimulus at the boundary compared to a stimulus lying within a category), and quantitatively (working on published experimental data on phoneme identification tasks).

\section{Introduction}
This paper addresses issues specific to the perception of categories (e.g. vowels, familiar faces, colors, ...), making a clear distinction between identifying a category (an element of a discrete set) and estimating a continuous parameter (such as a direction).
Categorization is long known to have an influence on perceptual judgments, as illustrated by many experiments based on discrimination and/or categorization tasks.
In particular, a perceptual phenomenon called categorical perception states that discrimination accuracy is higher at the boundary between categories than within a category \citep[see][for a review]{Harnad_1987}. 
This phenomenon has been much studied by psycholinguists in the case of phonemic categories -- 
languages differing by the number and the distribution of their phonemic categories, language acquisition indeed entails specific perceptual abilities \citep{Abramson_Lisker_1970,Werker_Tees_1984,Kuhl_etal_1992,Polka_Werker_1994}.
In addition to discriminability and categorization performances, many studies have measured reaction times. In the case of phoneme identification tasks 
(in all that follows, identification will denote identification of a category), it has been noted by \citet{Pisoni_Tash_1974} that ``reaction time is a positive function of uncertainty, increasing at the phonetic boundary where identification is least consistent and decreasing where identification is most consistent''. These authors thereby noted that ``identification time is slowest for the stimulus region where discrimination is best.'' 
Although this remark was formulated several decades ago, little attention has been given to the understanding
of the link between these two phenomena, discrimination and
identification time. Such understanding first requires to take into account the existence of a stimulus-dependent perceptual noise, second to determine how this affects decision making, and last to study how learning or adaptation jointly shapes  perceptual noise and reaction times. However, in previous models of categorization, discriminability is usually considered as a scale parameter, constant along a given relevant stimulus or psychological dimension, as in exemplar models \citep{Nosofsky_1986,Kruschke_1992}. In  \citet{Ashby_Maddox_1993},
the possibility of having a stimulus-dependent discriminability is for the first time considered, and later taken into account in the computation of reaction times
\citep{Ashby_Maddox_1994,Ashby_2000}.
Yet, to our knowledge, the information processing nature of both the  discriminability and its link with categorization has never been explored.
One of the main outcome of the present work 
is precisely to derive, from the hypothesis of optimal decoding, an analytical stimulus-dependent relationship
between mean reaction times and discrimination accuracy.\\

The above-mentioned psycholinguistic studies are particularly interesting for they exemplify how category learning affects both stages of neural processing, the encoding (the building of a neural representation of the stimuli) and the decoding (the reading-out of the categorical information and the decision-making process) ones. We assume that the former stage characterizes performances in a discrimination task, and that the latter is revealed by the measure of reaction times in an identification task.
A perceptual system that encodes categories aims at minimizing the probability of misclassifying incoming stimuli. It has to face two sources of uncertainty. The first one is independent of the neural code and lies in the intrinsic confusion between classes. For instance, vowels typically overlap in stimulus space. The second source of uncertainty comes from the noisy response of the neurons. 
In a previous work \citep{LBG_JPN_2008}, we showed how these two types of noise 
interact at the coding level. More precisely,
adopting a population coding scheme and making use of information theoretic tools,
we quantified the coding efficiency of a neural representation with respect to a set of categories by means of the mutual information (a measure of statistical dependency) between the set of categories and the neural activity. 
We showed that this information is essentially proportional to the ratio between two Fisher information values:
in the numerator a term that is independent of the neural code and that quantifies the overlap between categories in stimulus space; in the denominator a term that only depends on the neural code and that quantifies the sensitivity of the population code to small variations in the input space. 
An optimized code (resulting from either learning or evolutionary adaptation) is then realized by allocating more neuronal resources at the boundary between classes in order to have a greater Fisher information value of the neuronal population in this region, which implies a better sensitivity. In other words, if the code is optimized, discrimination is greater between categories than within a category, hence categorical perception.
In this previous work,
optimality is defined as maximization of the mutual information between the categories and the coding layer. No issue specific to the decoding stage was addressed there, even though maximizing the mutual information amounts to minimizing the probability of an ideal observer to misclassify an incoming stimulus (this property is formally given by Fano's inequality; see \citealt[\S 2.10]{Cover_Thomas_2006}, for a general statement, and \citealt[\S 2.2]{LBG_JPN_2008}, for a formulation of this inequality within the present context).\\

In the present work, we focus on the decoding stage, 
in particular on how it depends on the stimuli characteristics and on the efficiency of the coding stage. 
We show how the two types of noise (the two types of Fisher information values mentioned above) play a crucial role in the optimal decoding properties, hence in particular in shaping the reaction times. 
To derive our results, we work within a probabilistic framework, and consider a neural model based on a population encoding scheme as in \citet{LBG_JPN_2008}, and closely related 
to other neural models of categorization -- notably \citet{Kruschke_1992,Ashby_2000,Ashby_etal_2007} and \citet{Beck_etal_2008}. 
In particular, the model can be seen as an extended version of the covering model proposed by \citet{Kruschke_1992}, where the covering map is here interpreted as a neuronal population (with noisy activities), each neuron being specific of some region in the input space, and with the addition of a decision process based on a random walk dynamics. It can also be seen as a simplified version of the SPEED model of \citet{Ashby_etal_2007} -- leaving aside the learning issues not addressed here --, in a way which allows for analytical results. 
The chosen model  is precisely a compromise between biological plausibility and mathematical simplicity, allowing for analytical treatments. After a detailed presentation of the model in Section \ref{sec:model}, we then proceed in Section \ref{sec:results} with the two following main points. \\
$\bullet $ First point: optimal read-out. We study a decoding layer that provides an estimation of posterior probabilities.
We derive the theoretical properties of the optimal Bayesian decoder of the categorical information embedded in the coding layer. A crucial result is the derivation of a relationship between optimal decoding from a Bayesian point of view, and encoding efficiency as quantified by the mutual information between the neural activity and the categories. In particular, this relationship shows that maximizing information makes it possible to have a better estimate (in the sense that its variance is reduced) of the posterior probabilities giving the likelihood of a class knowing a stimulus in the transition regions between categories, which are the main sources of classification errors. 
Then, and quite importantly, we 
show that the neural parameters (tuning curves in the coding layer and synaptic weights for the decoding layer) of the considered architecture  can be adapted to provide the optimal estimator as output of the network.\\
$\bullet $ Second point: decision process. We consider the decision making mechanism as a diffusion model applied to the output of our network. First introduced in psychology \citep{Link_Heath_1975,Ratcliff_1978,Ratcliff_etal_1999}, diffusion models have been proposed as general models of decision-making, notably to account for reaction times. Roughly speaking, in the case of a two-alternative choice, diffusion models assume that a decision variable, that carries evidence accumulated in favor of one or the other choice, evolves stochastically over time until it reaches some threshold, leading to the decision. This type of models has more recently gained consideration in the field of neuroscience \citep{Smith_Ratcliff_2004,Gold_Shadlen_2007}.
The general theory of first passage times allows one to analytically express the mean reaction times during a category identification task. 
Although the mathematical foundations of this theory are general, most neuroscience applications  assume that the variance of the decision variable is independent of the presented stimulus \citep[see e.g.][]{Huk_Shadlen_2005}. Some models take into account the possibility of a stimulus dependent variance, thus
exhibiting a relationship between perceptual noise and reaction times \citep[see e.g.][]{Ashby_Maddox_1994,Ashby_2000}. However none of these works consider
how both bias and variance in the diffusion model depend on the stimulus when assuming 
optimal 
decoding. 
In the present paper, the dependency in the stimulus 
-- in both its categorical specificity and its encoding quality --
is crucial. For the considered architecture, building on the first point (optimal read-out) which studies the interplay between 
coding efficiency
and optimal decoding, 
we exhibit a quantitative link between reaction times and discrimination as a function of the stimulus.
With the aim of comparing the predictions of our model with behavioral data, we make a link between microscopic quantities (tuning curves of the neurons, synaptic weights) and macroscopic quantities (discrimination accuracy). The resulting formula makes it possible to model quantitatively mean reaction times obtained in a psycholinguistic experiment by \citet{Ylinen_etal_2005}:
our analysis allows to better analyze the difference in behavior between two groups, one for which one may expect that encoding has been efficiently adapted to the considered stimuli, and one for which this is not the case.\\

Finally, in Section~\ref{sec:discussion}, we put the emphasis on the analysis of the interplay between identification and discrimination as revealed by psycholinguistic studies on phonemic perception, and on the confrontation with neurophysiological data, and discuss the possible extensions of the model.

\section{Model}
\label{sec:model}

\subsection{Identification of categories: probabilistic framework}
\label{sec:general}

We consider $M$ categories, subscripted by $\mu=1,\ldots,M$, and characterized by a probability of occurrence $q_\mu$, so that $\sum_\mu q_\mu=1$, and a density distribution $P(x|\mu)$, where $x$ denotes the stimulus. For instance, $x$ might represents the voice onset time (VOT) dimension
in the case of stop consonants ($x \in \mathbb{R}$), 
or the two or three first formants in the case of vowels ($x \in \mathbb{R}^2$ or  $\mathbb{R}^3$). However, for simplicity, in all what follows we will assume that the stimulus space is unidimensional, that is $x \in \mathbb{R}$ (although our general theory is easily generalized to the multidimensional case). A stimulus $x$ elicits a response $\mathbf{r}=\{r_1,\ldots,r_N\}$ from a population of $N$ neurons that aims at encoding categorical information in a distributed fashion. The neural activity $\mathbf{r}$ depends on the class $\mu$ only through the sensory input $x$:
\begin{equation} 
P(\mathbf{r}| \mu) = \int  P(\mathbf{r}|x) P(x|\mu) dx  
\label{eq:prmu}
\end{equation} 
We restrict our analysis to the following conditions:
\begin{enumerate}
 \item for any neural activity $\mathbf{r}$ there is a uniquely defined
stimulus value $\hat{x}$ which maximizes the likelihood of $\mathbf{r}$ given $x$;
\item the system operates in a regime of high signal-to-noise ratio: $\hat{x}$
is a good approximation of $x$ (e.g. $N$ is large and $\hat{x}$ converges to $x$ as $N$ goes to infinity).
\end{enumerate} 
The large $N$ limit, which is the appropriate regime for modeling a population code, allows to have a high signal-to-noise ratio even with noisy individual neurons. From the mathematical point of view, it allows to obtain analytical results, with the interesting properties typically given by terms of order $1/N$ -- the first non trivial terms in the large $N$ limit.
Similar results would be obtained for a small number of cells, with low noise or in the large time limit, provided the mean firing rates are functions of $x$ allowing to get a good estimate of the stimulus value.\\

For what concerns the read-out, we will assume that, given a neural activity $\mathbf{r}$ in the coding layer,
the goal is to construct as neural output an estimator 
$g(\mu|\mathbf{r})$ 
of the posterior probability $P(\mu|x)$, where $x$ indicates the (true) stimulus that elicited the neural activity $\mathbf{r}$. 
For a given stimulus $x$ and a neural activity $\mathbf{r}$, the relevant quality criterion is given by the divergence (or improperly, the distance)
$\mathcal{C}(x,\mathbf{r})$
between the true probabilities $\{P(\mu|x), \mu=1,...,M\}$ and the estimator
$\{g(\mu|\mathbf{r}), \mu=1,...,M\}$, defined as the Kullback-Leibler divergence (or relative entropy) \citep{Cover_Thomas_2006}
\begin{equation}
\mathcal{C}(x,\mathbf{r}) \equiv \sum_{\mu=1}^{M} P(\mu|x) \ln \frac{P(\mu|x)}{g(\mu|\mathbf{r}) }
\label{eq:cost}
\end{equation}
Averaging over $\mathbf{r}$ given $x$, and then over $x$, 
the mean cost induced by the estimation can be written:
\begin{equation}
\mathcal{\overline{C}} = - \mathcal{H}(\mu|x) - \int dx \, p(x) \int d^N\mathbf{r}\, P(\mathbf{r}|x) \sum_\mu P(\mu|x) \ln g(\mu|\mathbf{r})
\label{eq:cm1_bis}
\end{equation}
where $\mathcal{H}(\mu|x) = - \int dx\,p(x) \sum_{\mu=1}^{M} P(\mu|x) \ln P(\mu|x)$ is the conditional entropy of $\mu$ given $x$.
In Section~\ref{sec:results}, we will study  the properties of the optimal estimator -- optimal in the sense that it minimizes the above cost function (\ref{eq:cm1_bis}) --,
and discuss its neural implementation.

Note that our hypothesis on the optimality criterion of the read-out is to be contrasted with other approaches, such as in \citet{Beck_etal_2008} modeling random dot discrimination task experiments. There, the discreteness of the classes is not  taken into account from the point of view of optimal information processing: the network makes its decision from the optimal estimation of a continuous variable (the global direction of the stimulus).

\subsection{Neural modeling}
\label{sec:neuralmodel}
\begin{figure}[!htb]
\begin{center}
\includegraphics[width=\linewidth]{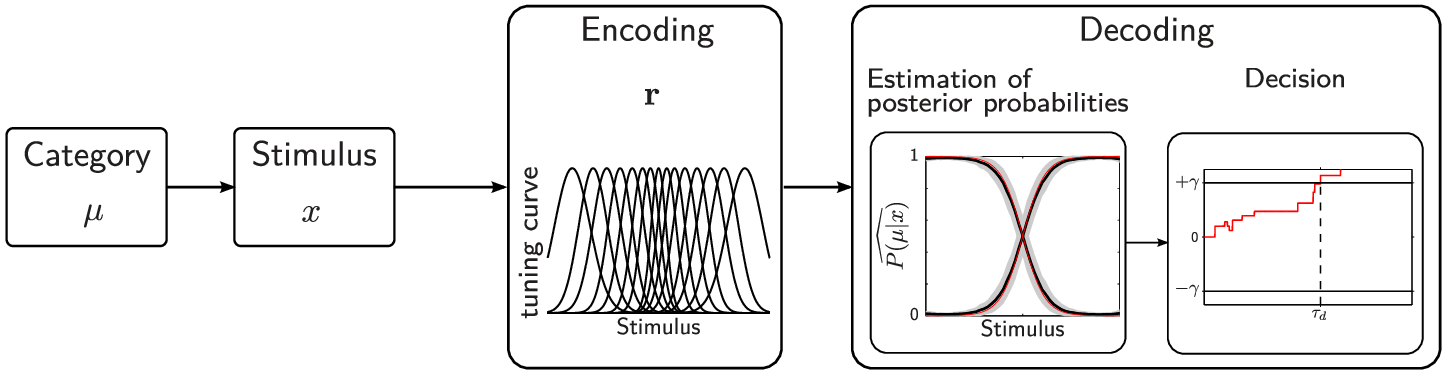}
\caption{\small
{Model architecture.}
Given a category $\mu$, a stimulus $x$ is produced according to some pdf $P(x|\mu)$. The stimulus is encoded by a large population of neurons with stimulus-specific tuning curves.
If the code has been optimized, more resources are allocated to the boundaries between categories in stimulus space, with tuning curves having steep slopes in the transition regions.
The information conveyed by the activity of this coding layer is extracted by the decoding layer. 
Thanks to an adaptation of the synaptic weights between the encoding and the decoding layer, the activity of the output cells (one per category) directly reflects category membership, by estimating the Bayesian posterior probabilities of the categories given a stimulus. The activities of these decoding units are the basis of the decision-making mechanism. In the case of two categories, the difference in activity between the two output cells acts as the decision variable through a diffusion process.
}
\label{fig:arch}
\end{center}
\end{figure}

We  now consider a plausible neural architecture. We assume a standard population coding scheme for the coding layer, followed by a decoding layer. This feedforward information processing is illustrated in Figure~\ref{fig:arch}.

\subsubsection{Population code}
\label{sec:popcod}
For the coding layer -- which we assume to characterize the perceptual level --, we consider an assembly of a large number $N$ of cells, with activities denoted $\mathbf{r}=\{r_1,\ldots,r_N\}$. Each cell $i$ is stimulus-selective, with a mean response characterized by a tuning curve $f_i(x)$
that peaks at its preferred stimulus $x_i$, and decreases according to some parameter $a_i$ (the width of the tuning curve). 
For simplicity we will assume that the  $r_i$'s are independent random variables given a stimulus $x$ (we will come back later on the important issue of correlations): 
\begin{equation}
P(\mathbf{r}| x) =\prod_{i=1}^N P_i(r_i|x)
\label{eq:model_factor}
\end{equation}
and we will assume Poisson statistics. Hence the mean number of spikes emitted during a time window $[0,\tau]$ 
and its variance 
are equal to $\tau f_i(x)$:
\begin{equation}
\left< r_i \right>_x =\tau f_i(x), \quad \left< (r_i)^2 \right>_x - \left< r_i \right>_x^2 =  \tau f_i(x)
\label{eq:fix2}
\end{equation}
where $\left< \,.\, \right>_x$ indicates the integration over $\mathbf{r}$ given $x$.
\subsubsection{Decoding layer: reading-out} 
\label{sec:decod}
The decoding layer aims at extracting categorical information from the neural population activity. 
This decoding layer features $M$ cells, each one connected to the $N$ neurons of the coding layer. 
The activity of decoding cell $\mu$ is given by a function $g(\mu|\mathbf{r,w})$, which will be interpreted as an estimator of the class likelihood, where the adaptable parameters are the synaptic weights $\mathbf{w} = \{w_{\mu i}, i =1,...,N, \mu =1,...,M\}$, $w_{\mu i}$ being the synaptic weight from the coding cell $i$ to the decoding cell $\mu$.
In order to constrain the neuronal activities so that the $g(\mu|\mathbf{r,w})$
can be interpreted as probabilities, 
we make the ad hoc (but standard) choice of a normalization with a softmax nonlinearity,
which constitutes a continuous generalization of the `winner-take-all' operation. For a given $\mu$, the $g(\mu|\mathbf{r,w})$ is then defined as:
\begin{equation}
g(\mu|\mathbf{r,w}) = \frac{\exp z_\mu}{\displaystyle \sum_{\nu=1}^M \exp z_\nu}
\label{eq:gmu}
\end{equation}
where, for every $\nu \in \{1,..., M\}$
\begin{equation}
z_\nu = \sum_{i=1}^N w_{\nu i} \, r_i.
\end{equation}

We now show that the estimator has a Gaussian distribution. Since the number 
$N$ of neurons is large, and the activities of the neurons being independent given an input $x$, according to the (Lyapunov's generalization of the) central limit theorem
$z_\mu$ is characterized by a Gaussian distribution 
with mean $\overline{z_\mu}$:
\begin{equation}
\overline{z_\mu}=  \tau \, \sum_i w_{\mu i} f_i(x)
\label{eq:mean}
\end{equation}
and variance $v(z_\mu)$:
\begin{equation}
v(z_\mu) =  \tau \, \sum_i w_{\mu i}^2 f_i(x).
\label{eq:var}
\end{equation}
Recall that $\tau f_i(x)$ represents the mean number of spikes emitted during a time window $[0,\tau]$. 
For large $N$ and (possibly) large observation time $\tau$, in order to have $\overline{z_\mu}$ of order $1$,
 the weights $w_{\mu i}$ must be of order $1/N\tau$, and then $v(z_\mu)$ is of order $1/N\tau$. Developing Eq.~(\ref{eq:gmu}) at first order in $1/N\tau$ shows that $g(\mu|\mathbf{r,w})$ also follows a Gaussian distribution, with
a mean $\overline{g_\mu}$, the average of $g(\mu|\mathbf{r,w})$ over $\mathbf{r}$
at a given value of $x$, of order $1$, and a variance $v(g_\mu)$, of order $1/N\tau$.
We will consider the expressions of these mean and variance in Section~\ref{sec:results}.

\subsubsection{Decision making from a diffusion process} 
\label{sec:rt}
The model is now completed by introducing the decision-making mechanism, which we present within the general framework of 
\textit{diffusion models}, which assume that information gets accumulated over time in favor of one or the other category until a threshold is reached, leading to the decision. 
This kind of models has been first introduced in psychology and provide good quantitative fits of psychophysical data \citep{Ratcliff_1978,Ratcliff_etal_1999}. Recently, it has found 
strong neurobiological supports \citep{Smith_Ratcliff_2004,Gold_Shadlen_2007}.\\
The analysis presented so far is valid for any number $M$ of categories. However, random walk or diffusion model only apply to two alternatives cases. In this part, and whenever appropriate, we thus restrict ourselves to the study of a two-category case. 
Generalizing the results to more than two categories would require considering other types of decision-making models such as accumulator models  \citep[see e.g.][]{Vickers_1970,Usher_McClelland_2001,Bogacz_Gurney_2007}.\\

As just said, a diffusion model
assumes that information gets accumulated over time in favor of one or the other category until it reaches a given threshold, leading to the decision \citep{Link_Heath_1975,Ratcliff_1978,Ratcliff_etal_1999}. This information is conveyed by a decision variable that favors one or the other category. When this variable, initially zero (whenever there is no preexisting bias), reaches the positive bound (notated $+\gamma$), the category corresponding to this bound (say category 2) is chosen. Conversely, when this variable reaches the negative bound (located in $-\gamma$), the other category (category 1 here) is chosen. As a consequence of the noise characterizing the temporal evolution of the decision variable, for the very same stimulus different trials might lead to different choices, and to different reaction times.
For the neural architecture studied here, the decision variable that we consider is the
difference between the output activities $z_{2,\tau}(\mathbf{r})$ and $z_{1,\tau}(\mathbf{r})$ -- that is the difference between the logarithm of the probabilities, $\log g(2|\mathbf{r,w}) /g(1|\mathbf{r,w})$. For a given time window $[0, \tau]$, this difference, notated $\alpha_{\tau}(\mathbf{r})$, is thus
\begin{equation}
\alpha_{\tau}(\mathbf{r}) = z_{2,\tau}(\mathbf{r}) - z_{1,\tau}(\mathbf{r})
= \sum_i (w_{2i}-w_{1i}) r_i
\label{eq:alpha_tau}
\end{equation}
where $r_i$ is the number of spikes emitted by neuron $i$ during the time window $[0, \tau]$.
If $\alpha_{\tau}(\mathbf{r})$ reaches the upper bound, $+\gamma$, (respectively the lower bound, $-\gamma$), the chosen decision is category 2 (resp. category 1).

As seen before, the number  $N$ of neurons being large, and the activities of the neurons being independent given an input $x$, one can make use of the central limit theorem and state that
$\alpha_{\tau}(\mathbf{r})$ is characterized by a Gaussian distribution,
and from (\ref{eq:alpha_tau}) 
and (\ref{eq:fix2}), one can write  its  mean $\overline{\alpha} (x)$ and variance $v_{\alpha} (x)$ as:
\begin{eqnarray}
\overline{\alpha} (x) &=& \tau \, \sum_i (w_{2i}-w_{1i}) f_i(x) \equiv \tau \overline{\alpha}^{0} (x)
\label{eq:meandiff} \\
v_{\alpha} (x) &=& \tau \, \sum_i (w_{2i}-w_{1i})^2 f_i(x) \equiv  \tau v_{\alpha}^{0} (x)
\label{eq:vardiff}
\end{eqnarray}
We have introduced the variables $\overline{\alpha}^{0}$ and $v_{\alpha}^{0}$ in order to make explicit the dependency in the time $\tau$.
Our diffusion process is thus characterized by the mean $\overline{\alpha} (x)$ and the variance $v_{\alpha}(x)$:
{\it both} depend on the stimulus $x$, not only the mean as often assumed in
the literature \citep[see e.g.][]{Huk_Shadlen_2005}.\\
Section~\ref{sec:results} will derive the mean time to reach one of the two bounds $+\gamma$ or $-\gamma$ (\textit{ie} the mean reaction time), and characterize the mean and variance of the decision variable in terms of both posterior probabilities of the categories and neural sensitivity of the coding layer.

\section{Results}
\label{sec:results}

This section develops and demonstrates the two main points of this paper, each of them consisting of two steps: 
(1a) characterization of the properties of the optimal decoder; this part, which might be found lengthy and technical, is however mandatory in order to understand how the efficiency of category identification (read-out accuracy measured by the appropriate Cram\'er-Rao
bound) is intrinsically linked to the efficiency of the encoding stage (measured by an information content); the results of this part are a crucial intermediate step upon which the
analysis of the reaction times is built;
(1b) within our neural model, neural implementation of the optimal decoder;
(2a) characterization of the mean reaction times as a function of the stimulus, assuming that the neural code has been optimized and (2b) interpretation of microscopic quantities (tuning curves of the neurons, synaptic weights) in terms of macroscopic quantities (discrimination accuracy).
The results are then illustrated by numerical simulations and confronted with experimental data.

\subsection{Optimal read-out: estimation of the posterior probabilities}

\subsubsection{Characterization of the optimal estimator}
Given a neural code, we here characterize, independently of the particular implementation of the decoding layer, the theoretical properties of the optimal estimator of the posterior probabilities of a category knowing a stimulus.\\
One can easily show that the estimator minimizing the cost function (\ref{eq:cm1_bis}) is
\begin{equation}
g(\mu|\mathbf{r}) = P(\mu|\mathbf{r}).
\end{equation}
One can expect the optimal estimator $P(\mu|\mathbf{r})$ to be unbiased and efficient in the large $N$ limit.
We show below that this is the case at leading order in $1/N$. In doing so, we derive from the Cram\'er-Rao bound an optimal bound for our cost function, and  provide an explicit link between the Bayes and the information theoretic approaches.\\

\noindent \textit{An unbiased estimator.}
Under the hypotheses presented above, one can show that -- up to a correction, hence a bias, of order $1/N$ that we will neglect in the following -- one has:
\begin{equation}
\int d^N \mathbf{r} \, P(\mu|\mathbf{r}) \,P(\mathbf{r}|x) = P(\mu|x) 
\label{eq:unbiased}
\end{equation}
Note that, because of the processing chain $\mu \rightarrow x \rightarrow \mathbf{r}$, the left-hand side of the above equation {\it is not} identically equal to $ P(\mu|x)$ (to be convinced, consider the zero signal-to-noise ratio case where the neural activity $\mathbf{r}$ does not depend on $x$). \\

\noindent \textit{Cram\'er-Rao bound.}
We here derive an optimal bound for the mean cost, Eq.(\ref{eq:cm1_bis}).
Let us consider an unbiased estimate $g(\mu|\mathbf{r})$
of the posterior probability $P(\mu|x)$ (that is $\int\,d^N\mathbf{r}\,g(\mu|\mathbf{r})  P(\mathbf{r}|x) = P(\mu|x)$). For such an estimate, the Cram\'er-Rao inequality writes \citep[see e.g.][\S 11.10]{Cover_Thomas_2006}:
\begin{equation}
\int d^N\mathbf{r}\,P(\mathbf{r}|x) \big(g(\mu|\mathbf{r})-P(\mu|x)\big)^2
\geq \frac{\big(P'(\mu|x)\big)^2}{F_\text{code}(x)}
\label{eq:cramer_rao}
\end{equation}
where $P'(\mu |x) \equiv \partial P(\mu |x)/\partial x$, and $F_{\text{code}}(x)$ is the Fisher information 
characterizing the sensitivity of $\mathbf{r}$ with respect to small variations of $x$:
\begin{equation}
F_{\text{code}}(x) = - \int \, d^N\mathbf{r}  \, \frac{\partial^2 \ln P(\mathbf{r}|x) }{\partial x^2} \, P(\mathbf{r}|x).
\label{eq:fisher_code}
\end{equation}
Now we rewrite the mean cost induced by the estimation, Eq.(\ref{eq:cm1_bis}), as:
\begin{equation}
\mathcal{\overline{C}} = -\int dx \, p(x) \int d^N\mathbf{r} \, P(\mathbf{r}|x) \sum_{\mu=1}^{M} P(\mu|x) 
\ln \frac{g(\mu|\mathbf{r})}{P(\mu|x)}
\label{eq:cm1_ter}
\end{equation}
For the typical values of $\mathbf{r}$ given a stimulus $x$, 
$g(\mu|\mathbf{r})$ has to be close to $P(\mu|x)$, so that:
\begin{equation}
\ln \frac{g(\mu|\mathbf{r})}{P(\mu|x)} = \ln \left(1+\frac{g(\mu|\mathbf{r})-P(\mu|x)}{P(\mu|x)}\right) 
\approx  \frac{g(\mu|\mathbf{r})-P(\mu|x)}{P(\mu|x)} - \frac{1}{2} \frac{\big(g(\mu|\mathbf{r})-P(\mu|x)\big)^2}{P(\mu|x)^2}
\end{equation}
Substituting this expansion within (\ref{eq:cm1_ter}) and using the fact that $g(\mu|\mathbf{r})$ is an unbiased estimate of $P(\mu|x)$, we can write 
\begin{equation}
\mathcal{\overline{C}} = \frac{1}{2} \int dx\,p(x) \sum_{\mu} \frac{1}{P(\mu|x)} \int d^N\mathbf{r}\,P(\mathbf{r}|x) \big(g(\mu|\mathbf{r})-P(\mu|x)\big)^2
\end{equation}
Hence, making use of the Cram\'er-Rao inequality~(\ref{eq:cramer_rao}), we get
that the mean cost satisfies:
\begin{equation}
\mathcal{\overline{C}} \geq \frac{1}{2} \int dx\,p(x) \frac{F_\text{cat}(x)}{F_\text{code}(x)}
\label{eq:ineqC}
\end{equation}
where $F_{\text{code}}(x)$ is the Fisher information (\ref{eq:fisher_code}),
and $F_\text{cat}(x)$ is the Fisher information that characterizes categorization uncertainty (which will be henceforth called the category-related Fisher information):
\begin{equation}
F_\text{cat}(x) = - \sum_{\mu}  \frac{\partial^2 \ln P(\mu|x)}{\partial x^2} \, P(\mu|x).
\label{eq:Fcat}
\end{equation}
Note that $F_\text{code}$ is of order $N$ and $F_\text{cat}$ of order $N^0=1$, so that the bound is of order $1/N$. Moreover, if the estimator has a bias of order $1/N$ (as this is the case below considering $P(\mu|\mathbf{r})$), one can show that the contribution of this bias to the Cram\'er-Rao bound is of order $1/N^2$, so that Eq.~(\ref{eq:ineqC}) remains valid.\\

\noindent \textit{An efficient estimator.}
If we now replace $g(\mu|\mathbf{r})$ in Eq.~(\ref{eq:cm1_bis}) by its optimal value $P(\mu|\mathbf{r})$, we get an interesting expression of the cost at the optimum, which is a difference between two mutual information values. Indeed, one can write
 $\mathcal{\overline{C}} =\mathcal{H}(\mu |\mathbf{r}) - \mathcal{H}(\mu | x)$, that is
\begin{equation}
\mathcal{\overline{C}} = I(\mu,x) - I(\mu,\mathbf{r})
\label{eq:Copt}
\end{equation}
where $I(\mu,x)$ is the mutual information between the categories $\mu$ and stimulus $x$,
\begin{equation}
I(\mu,x) \,=\, \sum_{\mu=1}^M q_{\mu}  \int dx \,P(x|\mu) \, \ln
\frac{P(x|\mu)}{p(x)} 
\end{equation}
and $I(\mu,\mathbf{r})$ the mutual information between $\mu$ and the neural activity $\mathbf{r}$
\begin{equation}
I(\mu,\mathbf{r}) \,=\, \sum_{\mu=1}^M q_{\mu}  \int d^N\mathbf{r} \,P(\mathbf{r}|\mu) \, \ln
\frac{P(\mathbf{r}|\mu)}{P(\mathbf{r})} 
\end{equation}

In \citet{LBG_JPN_2008}, we have shown that, in the large signal-to-noise ratio limit which we consider here, the difference $I(\mu,x) - I(\mu,\mathbf{r})$ which appears in the above equation (\ref{eq:Copt}) is given by:
\begin{equation}
I(\mu,x) - I(\mu,\mathbf{r}) = \frac{1}{2} \int dx\,p(x) \frac{F_\text{cat}(x)}{F_\text{code}(x)}
\label{eq:mi}
\end{equation}
that is precisely by the right hand side of the inequality (\ref{eq:ineqC}).
Hence, for the estimator $P(\mu|\mathbf{r})$, this inequality is an equality, which means that the Cram\'er-Rao bound is saturated.
The probability distribution $\{ P(\mu|\mathbf{r}), \mu=1,...,M\}$ 
is thus an estimator of $P(\mu|x)$ that is (asymptotically) unbiased and (asymptotically) efficient.\\

\noindent \textit{Information theoretic view point on Bayesian inference.}
En passant, we have thus shown that the decoding cost, for the optimal estimator, is directly related to the mutual information between the categories and the neural code. This result is in agreement with previous results in the field of statistical inference based on an information theoretic approach to Bayesian inference and neural coding \citep{Clarke_Barron_1990,Haussler_Opper_1995,Rissanen_1996,Herschkowitz_Nadal_1999,Bialek_etal_2001}: in words, the best estimator cannot do better than extracting the information that is conveyed by the available data/observations (here the neural activity) about the unknown parameter/stimulus (here the category). As a consequence, optimizing the code by maximizing the mutual information is also mandatory in order to optimize decoding. 

In addition, we have also obtained that the asymptotic expression (\ref{eq:mi}) of the mutual information, derived in 
\citet{LBG_JPN_2008}, has a nice interpretation since it comes from the Cram\'er-Rao bound. This is to relate to, and contrast with, the case of the coding (or estimation) of a continuous stimulus (or parameter) \citep{Clarke_Barron_1990,Rissanen_1996,Brunel_Nadal_1998}. If the aim of the considered neural system is to encode a continuous parameter $x$, e.g. an orientation, in the large signal-to-noise ratio limit the mutual information (between the neural code and the parameter) is essentially given by the logarithm of the Fisher information, $F_{code}(x)$, or more exactly stated by the logarithm of the bound of the Cram\'er-Rao inequality \citep{Brunel_Nadal_1998}. Here one gets that the mutual information (between the neural code and the category) is also expressed in term of the bound of the Cram\'er-Rao inequality, this bound being written for the estimation of the probability of the category (not of the category itself).\\

It follows from the previous results that an optimal strategy for the neural system consists in (1) applying the `infomax' principle to the coding layer; (2) building a decoding layer with $M$ output cells such that, from the neural activity $\mathbf{r}$ of the coding layer, the $\mu th$ output cell
has its activity precisely equal to the conditional probability $P(\mu|\mathbf{r})$.
One should note, however, that optimization of the decoding layer may be done for a given, not necessarily optimized, coding layer. 
It might be the case that the coding layer is used for different related tasks, and/or that the time scale for adaptation of the encoding is large, ensuring some long term stability or robustness despite 
the need to face various temporary tasks. In the case of linguistic data to be analyzed later,
the analysis will be consistent with the assumption that native speakers of a language have a well adapted neural representation  of their phonetic categories, whereas non native speakers do not.

\subsubsection{Network optimization}
In this section we consider the  optimization of the decoding layer through a learning procedure, this being done for a given coding layer (not necessarily optimized). The optimal estimator is
searched for within a class of probability distributions 
$\mathbf{g}(. |\mathbf{r},\mathbf{w}) \equiv \{g(\mu|\mathbf{r},\mathbf{w}), \mu=1,...,M\}$
that the neural system can implement,
$\mathbf{w}$ denoting the set of adaptable parameters (e.g. synaptic weights).\\

For an optimally adapted neural network, we thus expect $g(\mu|\mathbf{r},\mathbf{w})$ 
to have a distribution with mean $P(\mu|x)$ and with variance saturating the Cram\'er-Rao bound:
\begin{eqnarray}
\overline{g_\mu} = P(\mu|x) \label{eq:gmopt} \\
v(g_\mu) = (P'(\mu|x))^2/F_\text{code}(x)
\label{eq:gvopt}
\end{eqnarray}
For the considered neural model, we have seen that with $\overline{g_\mu}$ of order $1=(N\tau)^0$, for consistency one must have $v(g_\mu)$ of order $1/N\tau$. Since here $F_\text{code}(x)$ reads
\begin{equation}
F_\text{code}(x) = \tau \sum_i \frac{f_i^{'2}(x)}{f_i(x)},
\label{eq:fcode}
\end{equation}
this Fisher information is of order $N\tau$, hence the optimal variance given by (\ref{eq:gvopt}) is indeed of order $1/N\tau$.\\
As for the weights $\mathbf{w}$, although deriving general results sounds difficult, we expect the weights to be greater the further away the corresponding cell is to the category boundary: a cell `vote' should indeed be more important if it is more confident. One way to see that is to consider from Eqs.~(\ref{eq:gmu}) and (\ref{eq:gmopt}) that $\overline{z_\mu}=  \tau \, \sum_i w_{\mu i} f_i(x)$ behaves, up to a constant, as $\ln P(\mu|x)$; in the limit case of a continuum of cells with dirac delta function tuning curves, the weight function $w_{\mu}(x)$ is thus also proportional to the log of the posterior probability $P(\mu|x)$. In the following numerical illustration, the weights are indeed found to be greater within a category than between categories.\\

One may ask whether the chosen neural architecture allows to approximate efficiently the optimal solution. Actually general results on function approximation gives that a single `hidden layer' (here the coding layer) is enough in order to approximate any smooth enough function with an accuracy which can be as good as wanted with a large  enough number (here $N$) of `hidden units' (coding cells). In addition, making use of a very large number of coding/hidden units is in the line of the {\it support vector machine} (SVM) approach, which can be understood as projecting the inputs onto a large dimensional space, from which categorization becomes an easy task.
It is likely that many different learning algorithms, supervised or unsupervised, may be able to achieve 
the optimal solution. For illustrative purpose, in the following numerical simulations we will make use of a particular supervised learning strategy. 

\subsubsection{Illustration on two categories}
\label{sec:estim_num}
In this section, we illustrate our theory on the simplest example, that of two Gaussian categories. Recall that $x$ represents the relevant (continuous) physical space in which the stimulus lies. 
In the case of vowels, one may think of the space of formants. For comparison with specific empirical data, one may take as proxy for $x$ the 1-dimensional control parameter used in an experiment to make the stimulus changes continuously from one category to the other. For instance, in a face identification experiment, this dimension is defined by the morphed continuum between two different faces \citep[see e.g.][]{Beale_Keil_1995}. In the psycholinguistic study by \citet{Ylinen_etal_2005}, which will be studied in more depth in the following section, the control parameter is the vocalic duration. To fix ideas, consider the experimental study of \citet{McMurray_Spivey_2000}. In this experiment, subjects are presented with a continuum of $9$ stimuli, ranging from category \textipa{/ba/} to category \textipa{/pa/}, and whose voice onset time (VOT) values vary from $x_1=-50$ ms to $x_9=60$ ms. The task is to identify the category by clicking the corresponding button on a screen. Using an eye-tracking method, this behavioral study measures the time spent by subjects looking at the two buttons after hearing a given stimulus. Here we can consider the VOT as the relevant $x$-space. 

We assumed the two categories to be equiprobable, and each one characterized by a Gaussian distribution,
centered at $x^{\mu_1}=-2$ and $x^{\mu_2}=2$, with a width $a^{\mu_1}=a^{\mu_2}=1.5$ . These numbers are arbitrary and chosen for illustrative purpose only. 
For comparing the order of magnitudes
with the ones in the experiment described above, one unit of the $x$ space in the simulation corresponds to a difference in VOT of $13.75$ms (the spacing between two consecutive stimuli), with the categories centered at $x^{\mu_1}=-22.5$ms and $x^{\mu_2}=32.5$ms, and a width $a^{\mu_1}=a^{\mu_2}\sim 20.6$ms. 
We considered a neuronal population with $N=14$ coding cells. The activity $r_i$ of each neuron is given by a Poisson statistics with mean firing rate $f_i(x)$, corresponding to a bell-shaped tuning curve:
\begin{equation}
f_i(x) = f_{\min} + (f_{\max}-f_{\min})\, \exp \left( - \frac{(x-x_i)^2}{2a_i^2}\right)
\label{eq:bell}
\end{equation}
The preferred stimuli of the cells are equidistributed over the domain $[-6, 6]$ (which corresponds to VOTs in the range $[-77.5$ms, $87.5$ms$]$). 
The width and the minimal and maximal values of the tuning curves are the same for all the neurons $a_i=1.38$ ($\sim 19$ms),  
$f_{\min}=0.001$ and $f_{\max}=5$).

\begin{figure}[!htb]
\begin{center}
\includegraphics[width=.46\linewidth]{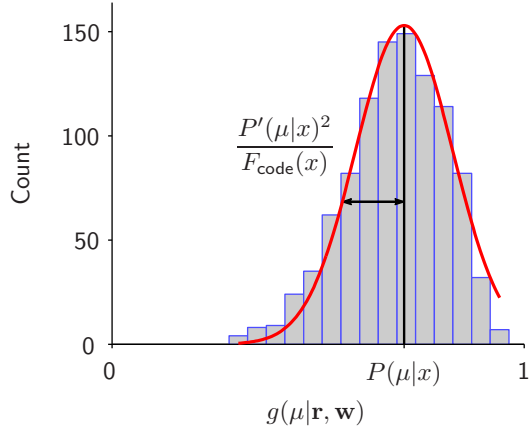}
\caption{\small
{Comparison between theoretical and numerical distribution of the posterior probability estimator.}
Histogram of the values of $g(\mu|\mathbf{r,w})$ ($\mu=1$) for $1000$ realizations of the neural activity $\mathbf{r}$ evoked by a stimulus close to the boundary between the two categories. 
In red, the theoretical curve: a normal distribution centered in $P(\mu|x)$ and with variance $P'(\mu|x)^2/F_{\text{code}}(x)$, predicting the values taken by an unbiased and efficient output.
}
\label{fig:dist_gmu}
\end{center}
\end{figure}

We ran a supervised learning phase in which a large number of stimuli $x$
are presented to the network along with their category label.
Following each presentation, the parameters $\mathbf{w}$ are updated in order to minimize the training cost function 
\begin{equation}
\mathcal{C}_t(x,\mathbf{r}) = \sum_{\mu=1}^M t_\mu(x) \ln \frac{t_\mu(x)}{g(\mu|\mathbf{r},\mathbf{w})}
\label{eq:t_cost}
\end{equation}
where $x$ is the presented stimulus, and the `teacher value' $t_\mu(x)$ is $1$ if the correct category is $\mu$, and $0$ otherwise. 
As shown in the Supporting Information,
through averaging over the presentation of a large number of stimuli, this cost becomes identical to the relative entropy between the true posterior probabilities and the output $g(\mu|\mathbf{r,w})$ (Eq.~\ref{eq:cm1_bis}).
Looking at the histogram of the values of $g(\mu|\mathbf{r,w})$ (following learning) for different realizations of 
the activity $\mathbf{r}$ evoked by a given stimulus $x$ (see Fig.~\ref{fig:dist_gmu}), we can notice the close proximity with the optimal theoretical curve given by the normal distribution centered in $P(\mu|x)$ and with variance $P'(\mu|x)^2/F_{\text{code}}(x)$. \\

\begin{figure}[!htb]
\begin{center}
\includegraphics[width=.97\linewidth]{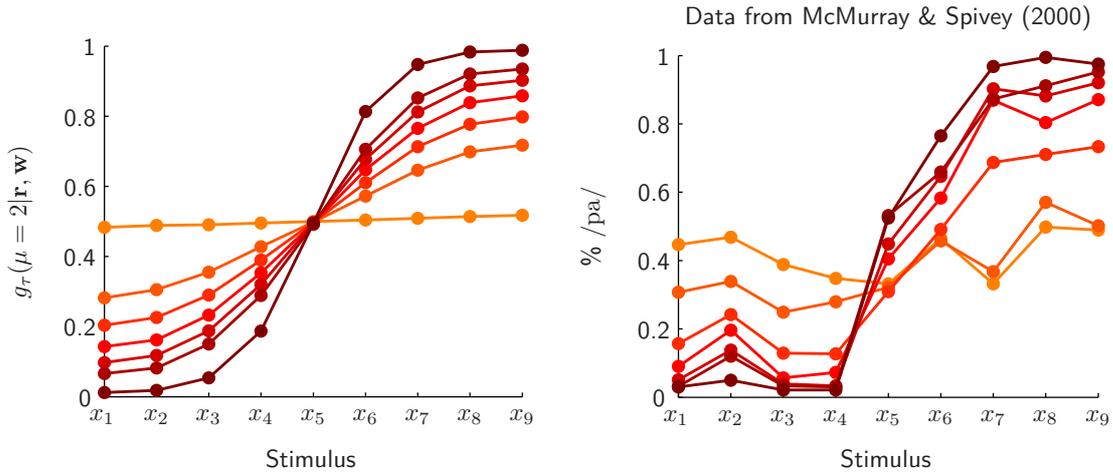}
\caption{\small
{Qualitative comparison of the temporal evolution of the decision between the model and as found in experimental data.}
(Left) Averaged temporal evolution of $g(\mu=2|\mathbf{r,w})$ along the continuum 
 $x_1, \ldots, x_9$.  
The increase in the length of the time window $[0, \tau]$ is indicated by a color gradient ranging from orange to dark red. (Right) Evolution of the proportion of looking time to the category \textipa{/pa/} vs the category \textipa{/ba/} for different stimuli whose voice onset time (VOT) values vary from $x_1=-50$ ms to $x_9=60$ ms \citep[data extracted from][]{McMurray_Spivey_2000}
}

\label{fig:sig_output}
\end{center}
\end{figure}

The temporal evolution of the output of the network reflects the accumulation of the categorical information extracted from the neuronal activity. The learning phase was performed on a time window $[0, \tau_{a}]$ so that $\tau_a f_{\max}$ represents the mean number of spikes emitted by cell $i$ during this time interval when the stimulus corresponds to its preferred stimulus. One can then look at the output $g(\mu|\mathbf{r,w})$ for different values of $\tau \in [0, \tau_{a}]$. 
Averaging over different realizations of this activity (1000 realizations in this numerical example), we finally get an estimate of the average value taken by the output $g(\mu|\mathbf{r,w})$ for each interval $[0, \tau]$.
Figure~\ref{fig:sig_output} (Left) shows the temporal evolution of the mean values of the output $g(\mu|\mathbf{r,w})$ for different stimuli along the continuum 
$x_1=-50$ms, $\ldots, x_9=60$ms 
(the curves getting redder and darker as $\tau$ increases). 
For comparison, Figure~\ref{fig:sig_output} (Right) shows the results from the
above-mentioned experimental study of \citet{McMurray_Spivey_2000}: one sees a gradual increase of categorical information, characterized by a sigmoid that expands over time, in qualitative compliance with our model.

\subsection{Reaction times}
\label{sec:res_rt}
This section analytically characterizes the mean reaction time following the identification of a category as a function of the stimulus presented, and shows how the analysis of the previous sections allows to specify, and understand the origin of, the parameters of the diffusion model.

\subsubsection{Mean reaction times}
\label{sec:mean_rt}
We want to express, in term of the threshold $\gamma$ and of the mean and variance of the diffusion process, the mean time $\overline{\tau_d}(x)$ to reach one of the two bounds. To do so we can apply to our model the general results on first passage times \citep{Wald_1947,Link_1992}. 
Applications of the theory of first passage time in the field of neuropsychology are presented in \citet{Shadlen_etal_2006}. 
The essential difference with these works is here the dependency of the variance in the stimulus.
The general theory on first passage time applied to our framework leads to the following equation:
\begin{equation}
\overline{\tau_d}(x) = \frac{\gamma^2}{v_{\alpha}^0 (x)} \,\Phi_d \left(\frac{\overline{\alpha}^0(x)\, \gamma}{v_{\alpha}^0 (x)} \right)
\label{eq:diff_gen}
\end{equation}
where
\begin{equation}
\Phi_d(y)  \equiv \frac{1}{y} \tanh(y)
\label{eq:Phi_d}
\end{equation}
with $\Phi_d(0)= \lim_{y\rightarrow 0}\Phi_d(y)=1$. 
One can get some insight on
the nature of this formula by considering an approximation 
which, although based on a two-lines argument, gives surprisingly good results.
First, to get rid of the sign, we consider the square  $\alpha_{\tau}(\mathbf{r})^2$ of the decision variable.
For a given time window $[0, \tau]$, 
we average this quantity over the realizations of the neuronal activity given a stimulus $x$. We then define (an approximation of) the mean reaction time $\overline{\tau_d}$ by the value of $\tau$ such that the average of $\alpha_{\overline{\tau_d}}(\mathbf{r})^2$ is equal to the square of the bound $\gamma$. In other words, we write
\begin{equation}
\left< (\alpha_{\overline{\tau_d}})^2 \right>_x = \gamma^2
\end{equation}
where $\left< \,.\, \right>_x$ indicates the integration over $\mathbf{r}$ given $x$.
Given the mean and variance, Eq. (\ref{eq:meandiff}) and (\ref{eq:vardiff}),
one gets a second-degree equation for $\overline{\tau_d}$, that is:
$ {\overline{\tau_d}}^2 \big(\overline{\alpha}^{0}(x)\big)^2 + \overline{\tau_d}\, v_{\alpha}^{0} (x) - \gamma^2 = 0$.
The positive root of this equation gives $\overline{\tau_d}$:
\begin{equation}
\overline{\tau_d}(x) = \frac{\gamma^2}{v_{\alpha}^0 (x)} \,\,\Phi_a \left(\frac{\overline{\alpha}^0(x)\, \gamma}{v_{\alpha}^0 (x)} \right)
\label{eq:taud_gen}
\end{equation}
where
\begin{equation}
\Phi_a(y) \equiv \frac{1}{2\,y^2} \left(- 1 + \sqrt{1 + 4\,y^2 } \right)
\label{eq:Phi_a}
\end{equation}
with $\Phi_a(0)= \lim_{y\rightarrow 0}\Phi_a(y)=1$.
Clearly the expressions~(\ref{eq:taud_gen}) and~(\ref{eq:diff_gen})
have the same structure. Despite the apparent dissimilarity between $\Phi_a$ and $\Phi_d$, these two functions have the same qualitative behavior as functions of their argument $y$, sharing the same asymptotic limits for both small and large values of $y$: both expressions for the mean reaction time give, for $| \overline{\alpha}^{0}| \gg v_{\alpha}^{0}(x)/\gamma$, $\overline{\tau_d}(x) \approx \frac{\gamma}{|\overline{\alpha}^{0}(x)|}$, 
and for $|\overline{\alpha}^{0}| \ll v_{\alpha}^{0}(x)/\gamma$,
$\overline{\tau_d}(x) \approx \frac{\gamma^2}{v_{\alpha}^{0}(x)}$.
Note that the similarity between our expression  (\ref{eq:taud_gen}) and the exact one (\ref{eq:diff_gen}) is remarkable since, in our argument, the notion of {\em first} passage is not even used.

\subsubsection{Macro interpretation of micro quantities}
We have seen that the mean and variance of the diffusion process result from the aggregation of information from the very large assembly of neurons in the coding layer. We now want to make use of our analysis on the optimal network, done in the previous sections, in order to give the expression of these mean and variance in terms of macroscopic quantities.\\

As we have shown, for the large $N$ limit considered here, the activity of the first output unit $g_{\tau}(1|\mathbf{r,w})$, 
is characterized by a Gaussian distribution. The 
mean $ \overline{g_{1,\tau}} $ and variance $v(g_{1,\tau})$ of this distribution can be easily determined:
\begin{equation}
\overline{g_{1,\tau}} = \frac{1}{1+\exp (\tau \overline{\alpha}^{0})}
\label{eq:d1:g1tau_m}
\end{equation}
and $ v(g_{1,\tau}) = \overline{g_{1,\tau}}^2 (1-\overline{g_{1,\tau}})^2 \, \tau^2 v_{\alpha}^{0} $, which can be rewritten as
\begin{equation}
 v(g_{1,\tau}) = \frac{{\overline{g_{1,\tau}}\,'}^2}{{\overline{\alpha}\,'^{\,0}}^2} \, \frac{1}{\tau} \, v_{\alpha}^{0}
\label{eq:d1:g1tau_v}
\end{equation}
where  $'$ denotes the derivative with respect to $x$, and we recall that
$\overline{\alpha}^{0}$ and $v_{\alpha}^{0}$ are the mean and variance of the diffusion process.

Now we have also just seen, Eq. (\ref{eq:gvopt}), that for the optimized network the mean $ \overline{g_{1,\tau}} $ and the variance $v(g_{1,\tau})$ are given by 
\begin{eqnarray}
\overline{g_{1,\tau_a}} &=& P(1|x)\\
v(g_{1,\tau_a}) &=& \frac{P'(1|x)^2}{\tau_a F_{\text{code}}^{0}(x)}
\end{eqnarray}
where $\tau_a$ is the integration time used during the learning phase, and $F_{\text{code}}^{0}(x)$
is the Fisher information rate specific to the neural code, so that if we observe the neural activity during a time window  $[0, \tau]$, the Fisher information of the neuronal population writes as $F_{\text{code}}(x) = \tau \, F_{\text{code}}^{0}(x)$.
For the present neural model, this Fisher information rate is given in term of the tuning curves by
\begin{equation}
F_{\text{code}}^{0}(x) = \sum_i \frac{(f_i'(x))^2}{f_i(x)}
\end{equation}

Making use of equations (\ref{eq:d1:g1tau_m}) and (\ref{eq:d1:g1tau_v}), we then get the mean and variance of the diffusion process in term of macro quantities: 
\begin{eqnarray}
\overline{\alpha}^{0} (x) &=& \frac{1}{\tau_a} \ln \frac{P(2|x)}{P(1|x)}
\label{eq:d1:alpha0}\\
v_{\alpha}^{0} (x) &=& \frac{{{\overline{\alpha}\,'^{\,0}}}^2}{F_{\text{code}}^{0}(x)}
\label{eq:d1:v0}
\end{eqnarray}
Given the expression of the bias (\ref{eq:d1:alpha0}), one can also write the variance as
\begin{equation}
v_{\alpha}^{0} (x) = \frac{1}{P(1|x)P(2|x)} \frac{F_{\text{cat}}(x)}{F_{\text{code}}^{0}(x)}
\end{equation}
where $F_{\text{cat}}$ is the category-related Fisher information, Eq.~(\ref{eq:Fcat}).
Note that
in accordance with the previous analyzes, $\overline{\alpha}^{0} (x)$ is of order $1/\tau_a$ and $v_{\alpha}^{0} (x)$ is of order $1/\tau_a^2$. \\

This analysis gives one of the main results of the present paper. It makes it possible to better understand the respective role of the mean $\overline{\alpha}^{0} (x)$ and the variance $v_{\alpha}^{0} (x)$ in the decision process. For a given stimulus $x$,  the diffusion bias $\overline{\alpha}^{0} (x)$ determines the mean direction taken by the decision variable towards one of the two bounds. This bias is 
given by the loglikelihood ratio favoring one hypothesis over another.
This is in agreement with previous works
on the Bayesian approach to decision making \citep[see e.g.][]{Gold_Shadlen_2007}, but note that here this is a result of the network optimization.
Within a category, $\overline{\alpha}^{0} (x)$, either negative or positive depending on the category, is characterized by a large value, which rapidly leads the decision variable to the correct corresponding bound. Conversely, at the boundary between categories, $\overline{\alpha}^{0} (x)$ is zero: the trajectory of the decision variable is then an unbiased random walk. The quantity $v_{\alpha}^{0} (x)$ determines the amplitude of the randomness in the trajectory of the decision variable. It is proportional to the ratio of the category-related Fisher information to the coding Fisher information. Recall that these Fisher information values give the sensitivity to small variations in the stimulus of, respectively, the category and the neuronal population.\\

{\em \noindent Application to Gaussian categories.}
If the categories are defined by Gaussian distribution with same variance, the quantity $\overline{\alpha}^{0} (x)$ is linear in $x$:
\begin{equation}
\overline{\alpha}^{0} (x) = b_0 (x-x_f)
\end{equation}
where $b_0$ is a scalar, and $x_f$ represents the boundary between categories, defined as $P(1|x_f)=P(2|x_f)$. In this case, $v_{\alpha}^{0} (x)$ simply writes:
\begin{equation}
v_{\alpha}^{0} (x) = \frac{b_0^2}{F_{\text{code}}^{0}(x)}
\end{equation}
Introducing the parameter $\beta \equiv \gamma/b_0$, the mean reaction time
takes a simpler expression:
\begin{equation}
\overline{\tau_d}(x) = \beta^2 F_{\text{code}}^{0}(x) \,\, \Phi\left(\beta\, F_{\text{code}}^{0}(x)\,(x-x_f)\right)
\label{eq:taud_gauss}
\end{equation}
where, for the exact expression (\ref{eq:diff_gen}),
$\Phi = \Phi_d$, Eq.~(\ref{eq:Phi_d}), 
and $\Phi = \Phi_a$, Eq.~(\ref{eq:Phi_a}), in the case of our approximation (\ref{eq:taud_gen}).\\
One can notice that Eq.~(\ref{eq:diff_gen}) and (\ref{eq:taud_gauss}) are (obviously) similar to those derived in previous models based on diffusion models. Notably, Eq.~(\ref{eq:taud_gauss}), corresponding to Gaussian categories that lead to linear decision bounds, is the same as in
\citet{Ashby_2000} (for identical absolute values of the negative and positive thresholds, and in the absence of `criterial noise' -- noise on the decision boundary). The key difference is in the interpretation of the parameters, 
here derived from the hypothesis of optimal decoding.
In particular, 
$F_{\text{code}}^{0}(x)$ is interpreted both in term of the discriminability measured in psychophysics, and in term of the neural sensitivity -- hence subject to adaptation. In  \citet{Ashby_2000}, in place of the Fisher information, the parameter which appears in the formula also characterizes the variance in the perception of the stimulus, but its characteristics are assumed independent of the categorization task.
In addition, in our result (\ref{eq:taud_gauss}), the constant  $b_0$ is analytically determined, in particular in terms of the posterior probabilities $P(\mu|x)$. This makes it possible to better predict or analyze the behavior of reaction times as a function of the structure of the categories. For instance, considering two Gaussian categories with equal variance, increasing the  variance of these distributions, which amounts to increase categorization uncertainty, results in longer reaction times, in a way that is quantified by our formula.\\\
We now apply the formula (\ref{eq:taud_gauss}) to data from a numerical simulation, and to experimental data available in the psycholinguistic literature.

\subsubsection{Numerical illustration}
We first test our theory with a numerical simulation on the simple case of two equiprobable Gaussian categories. The coding layer is composed of a (not so large) number of $N=10$ cells 
(see the Supporting Information for all the numerical details).
Given that we are interested in looking at the interplay between reaction times and discrimination, we here optimize both the coding layer and the decoding layer: the parameters of the tuning curves (width and location) in the coding layer are also optimized.

\begin{figure}[p]
\begin{center}
\includegraphics[width=.5\linewidth]{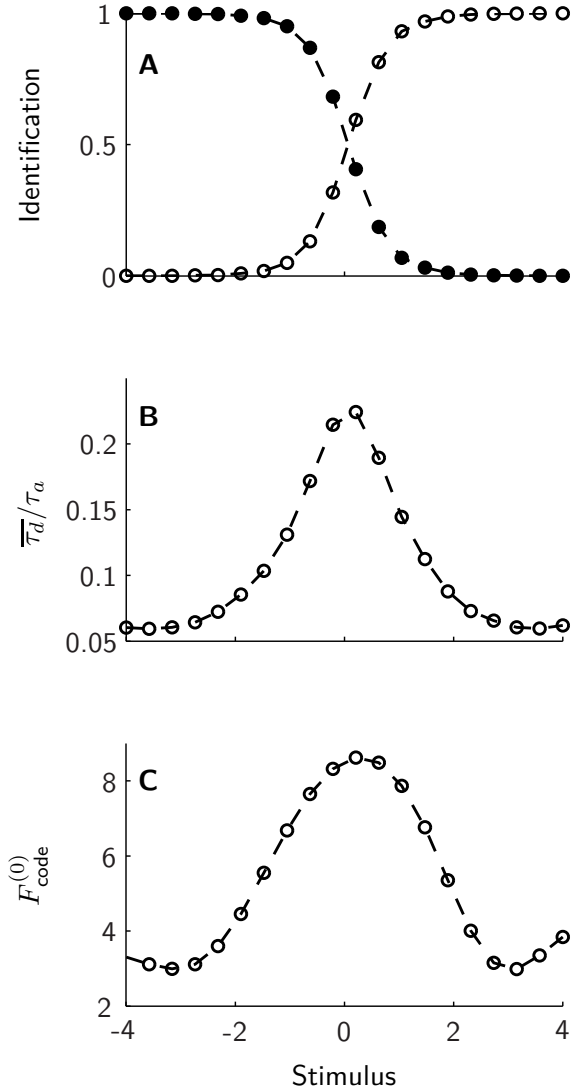} 
\end{center}
\caption{\small
{Perceptual consequences of category learning: results of the numerical simulation.}
(A) Mean identification function. (B) Mean reaction times. (C) Fisher information rate of the neuronal population (measure of perceptual sensitivity). These results qualitatively reproduce a classic situation of category learning, in particular in the case of phonemic perception \citep[see e.g.][Fig. 3]{Pisoni_Tash_1974}. Identification curves are characterized by an S-shape;  mean reaction times are longer at the boundary between categories than within category; discrimination accuracy (as quantified by Fisher information $F_{\text{code}}^{0}(x)$) is higher at the boundary between categories than within, \textit{ie} the neural population exhibits categorical perception.
} 
\label{fig:rt_num_data}
\end{figure}

Following learning, the behavior of the neural population, with respect to discrimination sensitivity and reaction times, qualitatively reproduces a classic situation of categorical perception, as summarized in Figure~\ref{fig:rt_num_data}.
Identification curves are characterized by an S-shape;  mean reaction times are longer at the boundary between categories than within category \cite[see e.g.][]{Pisoni_Tash_1974,Studdert-Kennedy_etal_1963}; discrimination accuracy (as quantified by Fisher information $F_{\text{code}}^{0}(x)$) is higher at the boundary between categories than within \cite[e.g.][]{Liberman_etal_1957,Repp_1984,Bornstein_Korda_1984,Goldstone_1994,Kuhl_Padden_1983}, which captures the so-called categorical perception phenomenon.

\begin{figure}[!htb]
\begin{center}
\includegraphics[width=\linewidth]{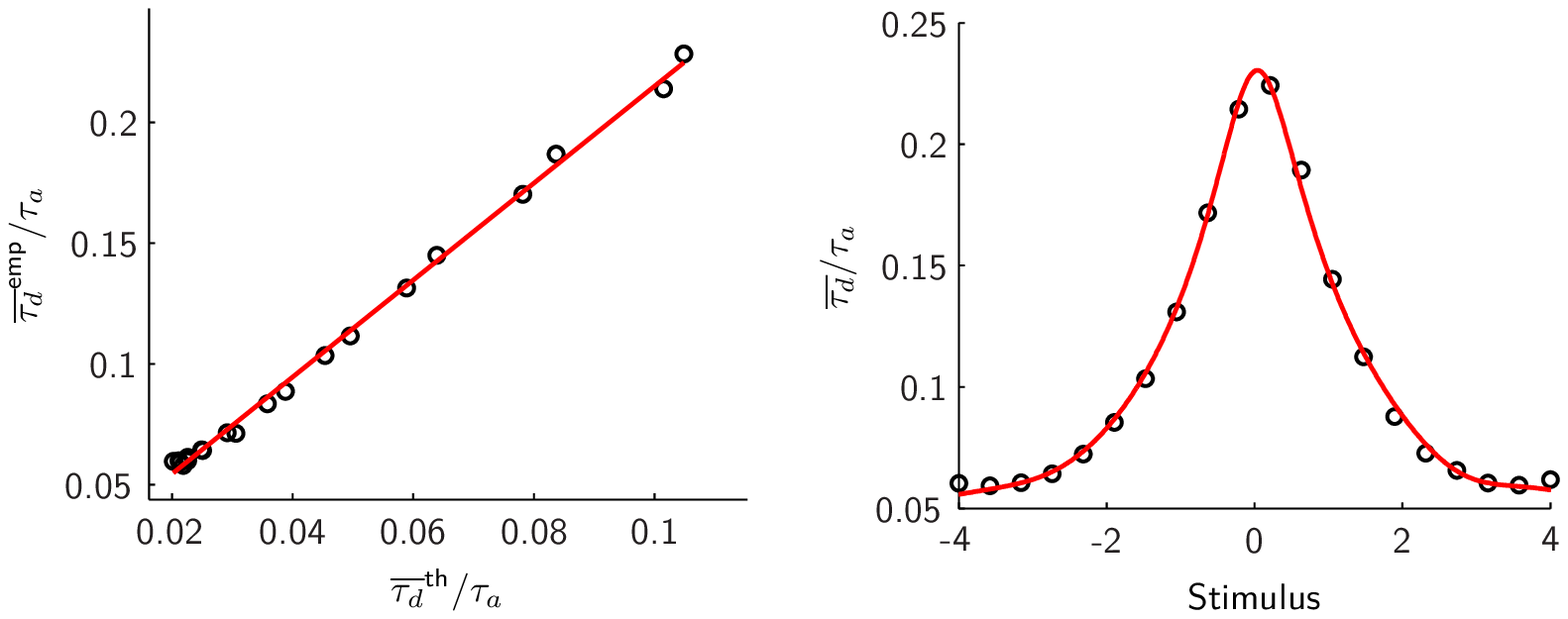}
\end{center}
\caption{\small
{\bf Reaction times: comparison between simulated data and theoretical prediction.} (Left) Mean reaction times $\tau_d^{\text{emp}}$ obtained by numerical simulation for the 20 stimuli spanning the considered continuum, as a function of the mean reaction times given by Eq.~(\ref{eq:taud_gauss}). The red line corresponds to the linear regression (correlation coefficient r=0.9986, p=1.7e-24).
(Right) Mean reaction times as a function of the stimulus presented. The open circles indicates the mean reaction times obtained by numerical stimulation, whereas the red line corresponds to the results derived from Eq.~(\ref{eq:taud_gauss}), (\ref{eq:Phi_a}).
} 
\label{fig:rt_num_pred}
\end{figure}

Figure~\ref{fig:rt_num_pred} (Left) compares the mean reaction times obtained in the numerical simulation with the ones predicted from formula (\ref{eq:taud_gauss}) and (\ref{eq:Phi_a}).
We can first emphasize the remarkable correspondence (up to a scaling factor) between the
simulated data and the data predicted by our equation, despite the fact that there is only $10$ cells in the coding layer. 
Using parameters of the linear regression extracted from Fig.~\ref{fig:rt_num_pred}, we can then reconstruct the mean reaction time for the whole continuum. This reconstructed mean reaction time is shown on Figure~\ref{fig:rt_num_pred} (Right, red line), together with the values obtained in the simulation (open circles). Here again, one can note the remarkable correspondence between the simulated and predicted values.
Note though that the values given by our formula (see the $x$-axis in Fig.~\ref{fig:rt_num_pred} (Left)) are smaller than the true values, hence the need in each case of rescaling the data in order to reconstruct the simulated reaction times. We attribute this bias to finite size and discretization effects.

\subsubsection{Modeling experimental data}
\label{sec:rt_comp}

\begin{figure}[!htb]
\begin{center}
\includegraphics[width=\linewidth]{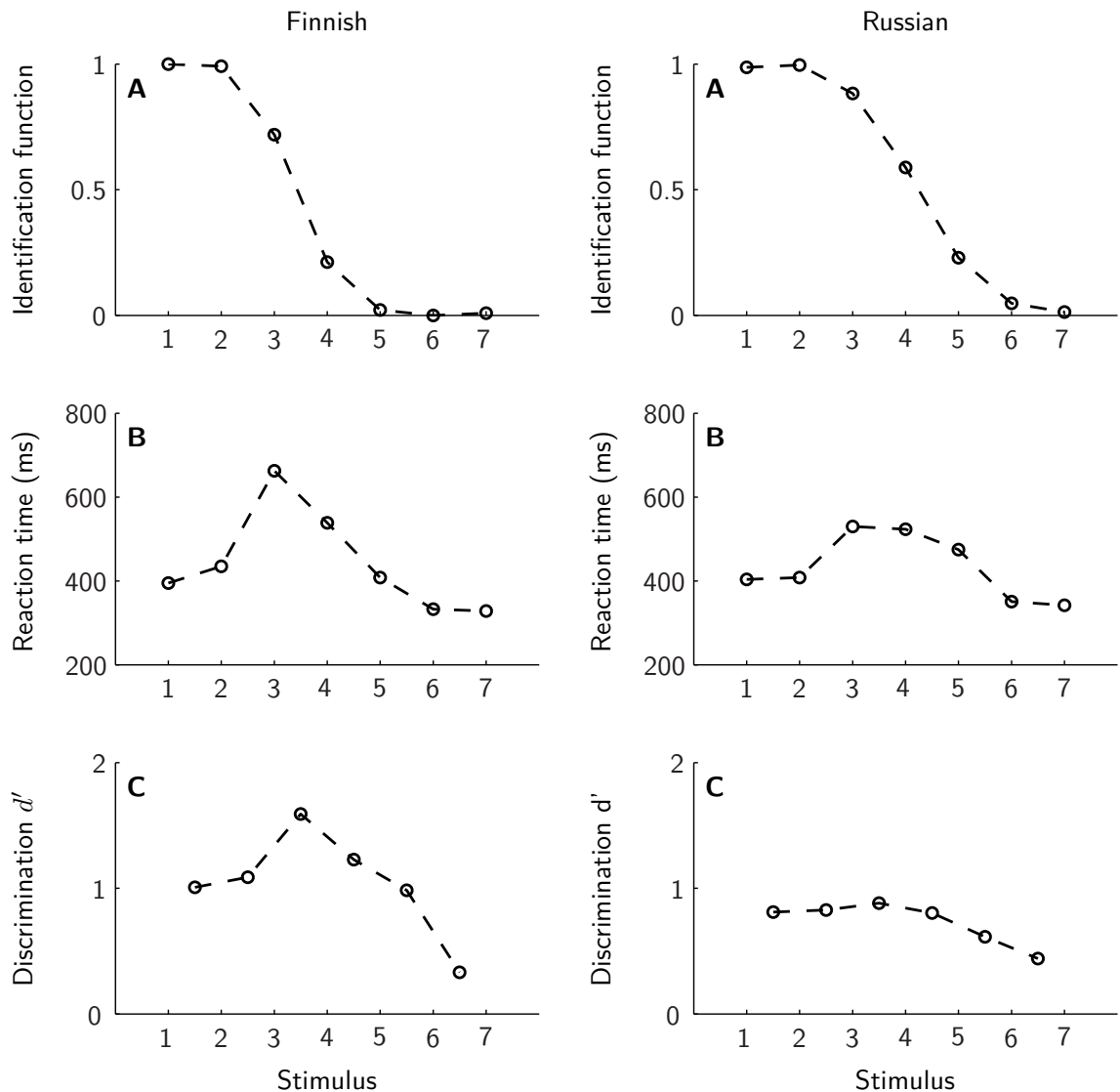}
\end{center}
\caption{\small
{Reproduction of the experimental data from \citet[][Experiment 2]{Ylinen_etal_2005}}. On the left, data corresponding to native speakers of Finnish; on the right,
those corresponding to Russian speakers. (A) Identification function. (B) Mean reaction times. (C) Perceptual distance ($d'$) between adjacent stimuli.
}
\label{fig:rt_xp_data}
\end{figure}

This section applies our theory to the modeling of mean reaction times obtained in the psycholinguistic study by \citet[][Experiment 2]{Ylinen_etal_2005}. This experimental study compares  the behavioral performances of two groups of subjects with respect to the perception of a phonological quantity based on duration. In this case, the two categories considered by the authors of this study are the two vowels \textipa{/u/} (short vowel) et \textipa{/u\textlengthmark/} (long vowel), the contrast being based on vocalic duration. For the first group of subjects (native speakers of Finnish), this contrast is phonemic, \textit{ie} these subjects have a distinct representation of the two categories. For the second group of subjects (Russians), the vocalic quantity is not contrastive. All the subjects were tested on a continuum of 7 stimuli.\\
A major interest for us here is that this study measures not only the reaction time during the identification of categories for each of the 7 stimuli along the continuum, but also the perceptual distance $d'$ between adjacent stimuli as well as reaction times during the discrimination phase (see Fig.~\ref{fig:rt_xp_data} for a reproduction of these data for the two groups of subjects).
These two latter sets of measurements make it possible to evaluate Fisher information rate for the whole continuum.\\

For a given stimulus $x$, the mean reaction time $\overline{\tau}(x)$ to categorize it is equal to the sum of the mean time $\tau_{nd}$ resulting from neural propagation and motor realization (independently of the decision), and the mean time $\overline{\tau_d}(x)$ characterizing the decision stage:
\begin{equation}
\overline{\tau}(x) = \tau_{nd} + \overline{\tau_d}(x)
\label{eq:tau}
\end{equation}
The mean time $\overline{\tau_d}(x)$ is given by formula (\ref{eq:taud_gauss}) (using $\Phi_a$ here), and depends on three free variables: $F_{\text{code}}^{0}(x)$, $x_f$, and $\beta$. We first determine the Fisher information rate $F_{\text{code}}^{0}(x)$ thanks to experimental measures of $d'$ and corresponding mean reaction times (measured during the discrimination task). The Fisher information rate $F_{\text{code}}^{0}(x)$ is linked to the perceptual distance $d'$  through \citep[see e.g.][]{Seung_Sompolinsky_1993}:
\begin{equation}
d'=|\delta x| \sqrt{F_{\text{code}}(x)}
\end{equation}
where, here, $\delta x = 1$. Moreover, as we have seen
\begin{equation}
F_{\text{code}}(x) = \overline{\tau_{d}}^{\text{discrim}}\,F_{\text{code}}^{0}(x)
\end{equation}
where $\overline{\tau_{d}}^{\text{discrim}}(x)$ corresponds to the mean reaction times during the discrimination task. For a given stimulus $x$, we compute the quantity $\overline{\tau_{d}}^{\text{discrim}}(x)$ thanks to the mean reaction times measured by the authors, equal to  
$\overline{\tau_{RT}}^{\text{discrim}}(x) = \overline{\tau_{nd}}^{\text{discrim}} + \overline{\tau_{d}}^{\text{discrim}}(x)$, where $\overline{\tau_{nd}}^{\text{discrim}}$, 
the mean time resulting from neural propagation and motor realization, is independent of the decision, and is set to $250$ms. Applying a piecewise cubic Hermite interpolation to the experimentally measured values, we obtain an estimation of $d'$ and of $\overline{\tau_{RT}}^{\text{discrim}}(x)$, and thus of $F_{\text{code}}^{0}(x)$, for all $x$ in the continuum.\\

\begin{figure}[!tb]
\begin{center}
\includegraphics[width=\linewidth]{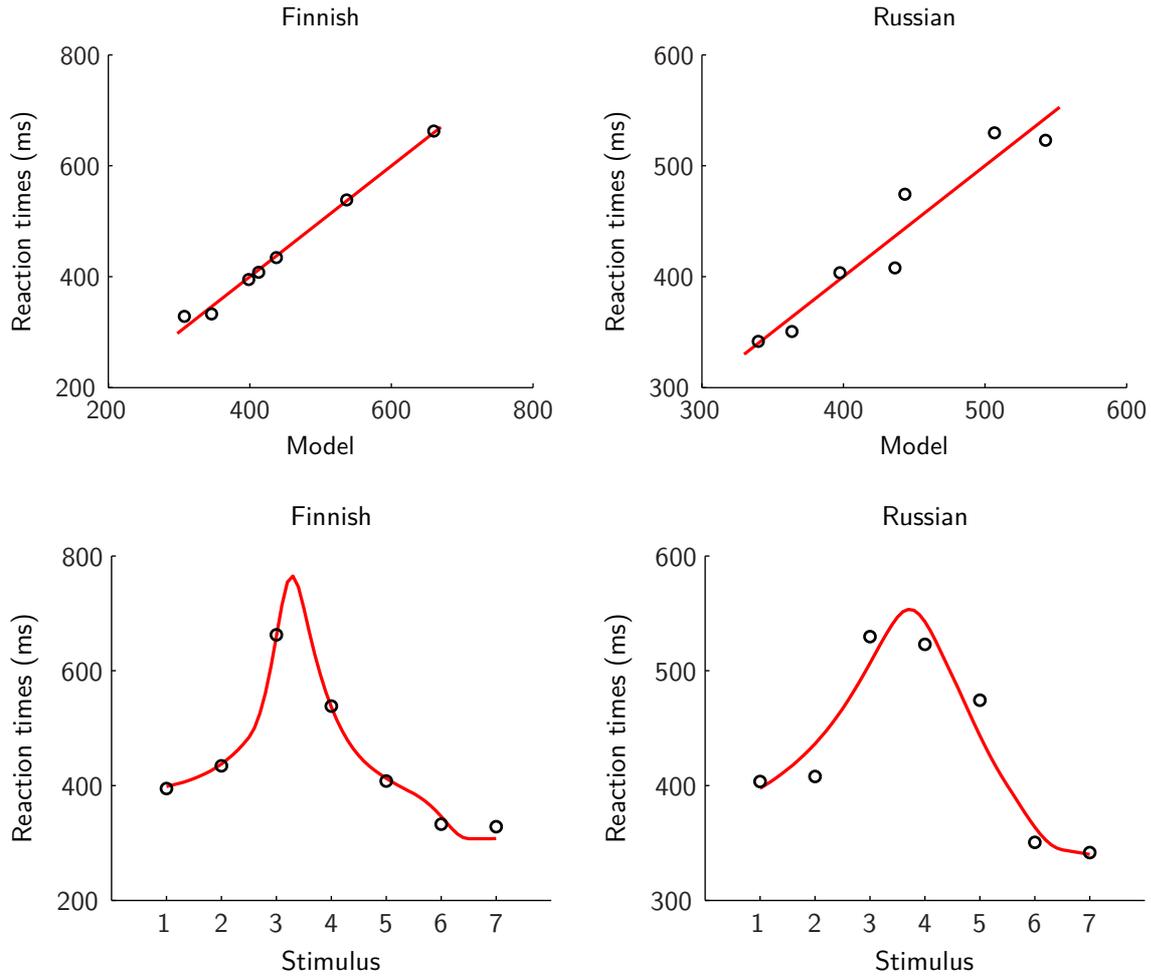}
\end{center}
\caption{\small
{Reaction times: comparison between experimental data and predictions from the model.}
On the left, the data corresponding to the native speakers of Finnish, on the right, those corresponding to the Russian group.
(Top)
Mean reaction times experimentally obtained by \citet{Ylinen_etal_2005} for the 7 stimuli spanning the considered continuum, as a function of the mean reaction times given by our formula, for the two groups of subjects. The red line corresponds to the $y=x$ line (r=0.996, p=1.7e-6 for the native speakers of Finnish, and r=0.959, p=6.5e-4 for the Russian group).
(Bottom)
Mean reaction times as a function of the stimulus presented, for the two groups of subjects. The open circles indicate the mean reaction times obtained in the experiment for each stimulus, whereas the red line corresponds to the model prediction.
} 
\label{fig:rt_xp_pred}
\end{figure}

Only three parameters are thus to be found in order to model the experimental data: $\tau_{nd}$,  $x_f$ and $\beta$.
For each group, these parameters are finally obtained by minimizing the least square error between experimental and predicted values. For the native speakers of Finnish, we get  $\tau_{nd}^{\text{fin}}=280$, $\beta^{\text{fin}}=339$ and $x_f^{\text{fin}}=3.11$ (r=0.996, p=1.7e-6), and for the Russian group, $\tau_{nd}^{\text{rus}}=278$, $\beta^{\text{rus}}=463$ and $x_f^{\text{rus}}=3.85$ (r=0.959, p=6.5e-4). 
Figure~\ref{fig:rt_xp_pred} compares mean reaction times experimentally obtained with the ones predicted by formula~(\ref{eq:taud_gauss}), optimized for each case. In the case of native speakers of Finnish  (Figure~\ref{fig:rt_xp_pred} (Left)), alignment between experimental data and prediction is almost perfect. In the case of the Russian group (Figure~\ref{fig:rt_xp_pred} (Right)), experimental data and predicted values line up remarkably well too.\\
Interestingly, the value of $\beta$ is found to be greater for the native speakers than for the non-native speakers. This parameter is equal to the ratio between $\gamma$, which is the decision threshold, and $b_0$, which quantifies the separation between the two categories. Thus, assuming one of the other parameter constant between groups, $\beta^{\text{fin}}<\beta^{\text{rus}}$ means that either the threshold is lower for the native speakers of Finnish than for the Russian group, or that the categories are more distinct for the natives. Both possibilities make sense here, given that we expect the native speakers to have a more accurate representation of the categories than the non-native speakers (the vocalic contrast used in this experiment being phonemic for the former group, but not for the latter).\\

It is worth noticing that only three free parameters are used here, thanks to the complete characterization of the $F_{\text{code}}^{0}(x)$ quantity using discrimination measurements. Other models of categorization response times do not allow for such a characterization, and would thus require more parameters.

\section{Discussion}
\label{sec:discussion}

\subsection{Interplay between identification and discrimination}
\label{sec:id_discrim}

The theory presented in this paper highlights the differences and relationships between 
identification and discrimination. The identification of categories is based on the output of the decoder, defined by the associative weights $\mathbf{w}$, whereas discrimination performance is determined at the level of the coding layer. In \citet{LBG_JPN_2008}, we showed that following category learning, neural optimization results in more neural resources allocated at the boundary between categories, with the aim of maximizing mutual information between neural activity and categories. Here, we have seen that optimization of the properties of the neuronal population entails a reduction of the uncertainty in the estimate of the posterior probabilities $P(\mu|x)$, which is particularly relevant in the transition regions between categories, and makes it possible to minimize classification errors.\\
This distinction is illustrated by the differences in the perception of a native speaker and of a second-language learner. A second-language learner has to associate sounds with new categories, \textit{ie} she has to build a decoder. After a learning phase, assuming no interference with existing category representation, this individual might then be able to correctly assign a label to the sounds she hears, thus presenting a response similar to the one produced by a native speaker. Nevertheless, this second-language learner will not necessarily exhibit a better discrimination at the boundary between categories.
In contrast, due to a more intensive experience and because the neural investment is behaviorally more relevant, a native speaker will typically exhibit a discrimination peak at the boundary between categories, which is a perceptual consequence of an optimized neural code (or `neural commitment', as \citet{Kuhl_2004} puts it.\\
This situation finds some experimental support in the study by \citet{Heeren_Schouten_2008} \citep[see also][]{Halle_etal_2004,Xu_etal_2006b}. Following an identification of categories, Dutch learners of Finnish exhibit a response curve similar to native speakers of Finnish, whereas naive Dutch speakers do not. Their discrimination curves however do not present a peak at the boundary, contrary to the native speakers. Following our analysis, more training and more language experience should lead a second-language learner to optimize her perceptual map, so as to better perceive fine variations at the class boundary. This is indeed the case: contrary to first- and second-year students, only third-year students present a discrimination peak at the category boundary.\\
Distinction between identification and discrimination is also reflected by reaction times. It is well known that reaction times follow some positive function of uncertainty: they are longer at the class boundary than within a category. As \citet{Pisoni_Tash_1974} note, the shape of the reaction times qualitatively follow the shape of the discrimination, typically greater between categories. We have seen though that this is not necessarily the case. Our result indeed show that longer reaction times at the boundary are inherent to the identification process, independently of a discrimination peak. We have exhibited yet a quantitative link between reaction times and discrimination accuracy (see Eq.~(\ref{eq:taud_gen})), showing that better discrimination implies longer reaction times (everything else being equal). We can thus predict that better discrimination at the boundary between categories results in a shape of reaction times that is sharper and with larger amplitude, which is supported by several experimental results \citep{Halle_etal_2004,Ylinen_etal_2005}.

\subsection{Neurophysiological data}
\label{sec:neuro}
In the studied model, a neural map encodes categorical information in a distributed fashion, so that if one only looks at a particular individual neuron, little information is conveyed, and the shape of the tuning curve does not reflect a categorical code. The influence of categorization on the neuronal properties has to be evaluated at the population scale. Conversely, the output cells, involved in the decision process, code in a more direct way for the categories. Their activities indeed follow the posterior probabilities related to the categories: a given cell responds similarly within a category and sharply differently between. This situation finds biological support in recent neurophysiological studies.\\
In particular, in the case of the visual system, the inferotemporal cortex (IT) encodes information in a distributed way, with  a population coding strategy, 
and feeds downstream prefrontal (PFC) regions characterized by more categorical responses.
Several studies have shown 
that category learning modify the neuronal properties of the IT population \citep{Sigala_Logothetis_2002,DeBaene_etal_2008,Kriegeskorte_etal_2008,OpdeBeeck_etal_2008b}.
In their study on the influence of categorization in the inferotemporal and prefrontal cortices, \citet{Freedman_etal_2003} conclude that there is (almost) no categorical information in the inferotemporal cortex, whereas cells in the prefrontal do show categorical specificity.
These arguments are based on a measure of categorical selectivity at the single neuron scale, which might potentially overshadow information collectively conveyed by the whole population. Several studies have yet shown an influence of category learning on neuronal properties in the inferotemporal cortex, and insist on the distributed coding strategy employed in the inferotemporal and the more individual code in the prefrontal \citep{OpdeBeeck_etal_2008b,Meyers_etal_2008,Kriegeskorte_etal_2008}.
Similarly, the MT (middle temporal) region,
thought to play a major role in the perception of motion and in the guidance eye movements,
is well modeled as a large population of direction-specific cells.
\\ Located downstream of the inferotemporal cortex, the prefrontal cortex is known as a site
for superior cognitive functions, notably decision-making. Several studies show that neurons in the prefrontal cortex have an activity that more directly reflects category membership, and which is not much affected by the physical properties of the stimulus itself \citep{Freedman_etal_2001,Freedman_etal_2002}. These neurons have typically a step-like tuning curve (or its continuous counterpart, an S shape), and exhibit a strong categorical selectivity at the individual level. We can also evoke here the existence of neurons that responds categorically following category learning, in both auditory cortex \citep{Ohl_etal_2001,Prather_etal_2009} and primary motor cortex \citep{Salinas_Romo_1998a}.\\
Concerning the decision mechanism and reaction times, several studies published in the past decade have brought quantitative support to the kind of diffusion model we used here, for which neuronal activity represents accumulation of evidence in favor of one or the other possible choice until it reaches a given bound, entailing the decision \citep{Kim_Shadlen_1999,Shadlen_Newsome_2001,Heekeren_etal_2004,Huk_Shadlen_2005,Smith_Ratcliff_2004,Gold_Shadlen_2007}. In the case of random dot discrimination task experiments where the decision is made through eye movements, the LIP (lateral intraparietal) region, which receives inputs from MT, has been identified as the locus of such decision mechanisms \citep{Shadlen_Newsome_2001}.\\
In our modeling, we have assumed uncorrelated cells (conditional to the stimulus). For the coding stage, the main results hold or are easily generalized \citep{LBG_JPN_2008,LBG_These}, whenever the 
noise correlations preserve the scaling of the Fisher information with the size of the neural assembly ($F_{code}(x) \sim N$) -- which is known to be the case for a large family of correlations \citep{Abbott_Dayan_1999,Yoon_Sompolinsky_1999}.
However, the hypothesis of uncorrelated cells plays an important role for the decoding layer, for which the results explicitly need that the output cells sum independent random variables.
Experimental results in favor of diffusion models are actually easily understood if this is the case. However some experimental works strongly suggest that important correlations exist in the coding layer \citep{Zohary_etal_1994}. To conciliate such results with the observed activities in the decoding areas, some authors proposed that the cells might sum a small number of well chosen cells in the coding layer \citep{Zohary_etal_1994,Britten_etal_1992}. We have seen that, in the numerical simulations, the results obtained with a rather small number of independent cells are already in good agreement with the analytical results assuming a large number of cells.
An alternative but related scenario is to assume that the correlations do decrease sufficiently fast with the difference in preferred stimuli, so that the effective number of independent cells seen by the decoding layer is of order $N/R$, where $R$ is the typical scale of the correlations. Provided $N\gg R$, one may expect the results presented here to apply as well. In addition, the existence of strong correlations has recently been challenged \citep{Renart_etal_2010,Ecker_etal_2010}, from analyses based on both theoretical and experimental approaches. Such controversial issue needs to be resolved by new experiments, and specific studies of optimal decoding with correlations remain to be done.\\
In any case, it thus appears 
that for different modalities and categorization tasks, the same global scheme is found: a distributed encoding with a large population of feature-specific cells, a read-out layer and a decision mechanism -- with a diffusion or accumulator mechanism. 
In \citet{LBG_JPN_2008}, we discuss the relevance of our approach to the modeling of, e.g., the IT neural assemblies as corresponding to the coding neural cells. 
Here our main results concern the decoding layer (e.g. PFC, LIP), and link this decoding layer to the coding stage. In particular, from our theory, one should find that both the bias and variance of the random walk process have a dependency in the stimulus. More precisely,
these parameters 
should be related to the class probabilities (given the stimulus) and to the sensitivity of the neural code with respect to the stimulus. \\

\subsection{Concluding remarks}
When dealing with a difficult categorization task, the brain has to face two independent sources of uncertainty: categorization uncertainty and neuronal uncertainty. The latter stems from neuronal noise, whereas the former is intrinsic to the category structure in stimulus space: categories like phonemes or colors typically overlap, so that a given stimulus might belong to different categories.  Here, we propose a general neural theory of category coding, in which these two sources of uncertainty are quantified by means of information theoretic tools. We analytically show how these two quantities combine at both coding and decoding stages of the information process. Considering optimal representations, we derive formulae which  capture different psychophysical consequences of category learning -- namely, a better discrimination between categories, and longer reaction times to identify the category of a stimulus lying at the category boundary. Finally, we analytically relate microscopic quantities (neural properties) to macroscopic quantities that are behaviorally measurable (discrimination accuracy): this allows us to model experimental data, in the present work taken from the psycholinguistic literature. A major contribution of this work is thus to exhibit, in both quantitative and qualitative terms, the interplay between discrimination and identification,
 thanks to a global approach which links
the `top-down' one -- the ideal observer approach where one compares the behavioral performance to the optimal ones --, with the `bottom-up' one -- the building of a neural 
code starting from the stimulus space.\\
The stimulus structure is here formalized within a probabilistic framework, and we considered a neural architecture aiming at extracting the
categorical information. The stimulus is encoded by a large population of stimulus-specific neurons, and the decoding  is achieved by a layer  of category-specific cells.
We have shown that the output of these cells can estimate the posterior probabilities 
giving the likelihood of the classes 
knowing the stimulus 
(in the simulations we considered a particular supervised learning scheme, but one can expect that others, supervised or unsupervised, can achieve the same results). Minimizing the Kullback-Leibler distance between the true probabilities and the output of the network leads not only to build an unbiased and efficient estimate of these probabilities but also, if the properties of the neuronal population are optimized, to maximize the mutual information between the activity of this neuronal population and the categories. Within such context, allocating more neural resources at the boundary between classes \citep{LBG_JPN_2008} makes it possible to minimize classification errors at the boundaries, thanks to a better estimation of posterior probabilities.
We have restricted the analysis to the case of a one-dimensional input: the present theory can easily be generalized to multidimensional inputs, as done in \citet{LBG_JPN_2008}.
The issue of correlations in the coding layer and its impact on decoding deserves more studies, as discussed in the above \ref{sec:neuro} section.\\
As explained previously the presented neural model share with others the same skeleton, but is simple enough to allow for analytical results.
The latter quantify the efficiency in the categorization task, and give the best possible performances that can be achieved through learning -- the specific issue of learning being not addressed here.
Despite this (relative) mathematical simplicity, the model preserves a strong biological plausibility.
The coding/decoding architecture
receives support from several recent experimental results in neurophysiology. A neuronal population encodes categorical information in a distributed fashion and then feeds downstream regions that use this information to realize higher cognitive functions, such as decision-making.
In the particular case of the visual system, this situation corresponds respectively to the IT and PFC (as found in visual object categorization tasks), as well as to the MT and LIP regions (as found in random dots experiments). At the level of the inferotemporal cortex, categorical information is distributed among the whole neuronal population, so that each neuron taken individually is not category specific. Conversely, in the prefrontal cortex, category membership is more explicitly represented at the level of a single cell, where information gets accumulated over time. This decisional process is here modeled by a diffusion model: a random variable, the difference in output activities, 
evolves over time until it reaches a certain bound, positive or negative, leading to the corresponding response. We have analytically characterized the mean reaction time necessary to the establishment of the decision, relating it to both a quantity that measures the degree of membership of the stimulus to one or the other category, and to the Fisher information quantifying the perceptual sensitivity in a discrimination task. We have shown that the formula we derived account for experimental data obtained in the psycholinguistics literature \citep{Ylinen_etal_2005}. This comparison is however based on data that involve a small number of stimuli and, more importantly, are only averages of the performances of a group of subjects. In order to test our model more precisely, future research should gather detailed individual behavioral data on discrimination accuracy and reaction times. Experiments on animal could provide the same type of data together with measurements of neural responses, at both the encoding and decoding level, so as to test the interplay between these two stages as we have discussed in this paper. Experiments should focus on the smooth transition between categories which, in view of our analysis, is the most relevant region to reveal both the sensitivity of neural code and the related shape of the reaction times.
\\Finally, we mention that within our framework the modeling of random dot experiments requires to consider the  extension to a time-fluctuating multi-dimensional stimulus -- in the vein of \citet{Ashby_2000} or \citet{Beck_etal_2008}. More importantly, it requires a specific analysis: as the level of coherence changes, almost every quantity (both Fisher information values, the bias and the variance of the diffusion process) is changed.
We leave to further work the study of the resulting dependency of reaction times in the coherence.

\section*{Acknowledgments}
This paper has benefited from critical readings by several colleagues. We especially thank Dr. F. G. Ashby
for pointing out to us important references.

\clearpage
\newpage
\appendix

\renewcommand{\thefigure}{\thesection.\arabic{figure}}
\renewcommand{\thetable}{\thesection.\arabic{table}}

\section{Network optimization: supervised learning scheme}
\label{sec:sup_learn}
\setcounter{equation}{0}

For the numerical simulations, we made used of a supervised learning scheme which we present here, proving  that, in the asymptotic limit of a very large training set, the chosen cost function
gives the cost $\overline{\mathcal{C}}$ considered in the theoretical analysis.

During learning, stimuli are presented sequentially, along with their category label.
For a given stimulus $x$, the output $g(\mu|\mathbf{r},\mathbf{w})$ is compared with the desired binary output given by indicator function $t_\mu(x)$ (for \textit{\textbf{t}eacher}), defined as follows:
\begin{equation}
t_\mu(x) = \left\{ \begin{array}{ll} 
		  1 & \text{if } x \in \mu\\
		  0 & \text{otherwise}
\end{array} \right.
\end{equation} 
where $x\in\mu$ means that stimulus $x$ belongs to the category labeled $\mu$. 
The distance between the output $g(\mu|\mathbf{r},\mathbf{w})$ and the teacher value $t_\mu(x)$, 
is measured by the following training cost function:
\begin{equation}
\mathcal{C}_t(x,\mathbf{r}) \equiv \sum_{\mu=1}^M t_\mu(x) \ln \frac{t_\mu(x)}{g(\mu|\mathbf{r},\mathbf{w})}
\label{eq_ap:t_cost}
\end{equation}
Its average over all the realizations of the neural activity $\mathbf{r}$ is given by:
\begin{equation}
\mathcal{C}_t(x) = \int d^N\mathbf{r}\, P(\mathbf{r}|x) \sum_{\mu=1}^M t_\mu(x) \ln \frac{t_\mu(x)}{g(\mu|\mathbf{r},\mathbf{w})}
\label{eq_ap:avg_cost}
\end{equation}

Let us now show that a large number of stimulus presentations during the learning phase leads to estimate posterior probabilities \citep[in a way similar to the one presented in][]{Duda_etal_2001}.\\
After $n$ stimulus presentations, the mean cost function becomes:
\begin{eqnarray}
\frac{1}{n} \sum_x \mathcal{C}_t(x) &=& \frac{1}{n} \int d^N\mathbf{r}\, \sum_x P(\mathbf{r}|x) \sum_{\mu} t_\mu(x) \ln \frac{t_\mu(x)}{g(\mu|\mathbf{r},\mathbf{w})}\\
&=& - \frac{1}{n} \int d^N\mathbf{r}\,\sum_\mu \sum_{x\in\mu} P(\mathbf{r}|x) \ln g(\mu|\mathbf{r},\mathbf{w})\\
&=& - \sum_\mu \int d^N\mathbf{r}\, \frac{n_\mu}{n} \frac{1}{n_\mu} \sum_{x\in\mu} P(\mathbf{r}|x) \ln g(\mu|\mathbf{r},\mathbf{w})
\end{eqnarray}
where $n_\mu$ is the number of stimuli labeled $\mu$ among the $n$ stimuli that were presented to the network.\\
For a very large number of stimuli, the mean cost $\overline{\mathcal{C}_t}$ then writes:
\begin{equation}
\overline{\mathcal{C}_t} \equiv \lim_{n\rightarrow \infty} \, \frac{1}{n} \sum_x \mathcal{C}_t(x)
= - \sum_\mu \int d^N\mathbf{r}\, q_\mu \int dx\, P(x|\mu) P(\mathbf{r}|x) \ln g(\mu|\mathbf{r},\mathbf{w})
\end{equation}
hence, given that $\int dx\, P(x|\mu) P(\mathbf{r}|x) = P(\mathbf{r}|\mu)$, and that, according to Bayes rules $q_\mu P(\mathbf{r}|\mu) = P(\mathbf{r}) P(\mu|\mathbf{r})$, we get
\begin{equation}
\overline{\mathcal{C}_t} = - \int d^N\mathbf{r}\, P(\mathbf{r}) \sum_\mu P(\mu|\mathbf{r}) \ln g(\mu|\mathbf{r},\mathbf{w})
\label{eq_ap:cost_learning}
\end{equation}
This is the same as $\overline{\mathcal{C}}$ 
except for a constant additive term (the entropy $\mathcal{H}(\mu|x)$), implying that minimization of the cost leads to estimate the posterior probabilities, as desired.

In the numerical illustrations, learning is done through a gradient descent algorithm \citep{Rumelhart_etal_1986} aiming at minimizing the cost function (\ref{eq_ap:t_cost}),
with the presentation to the network of 30000 stimuli along with their category label.

\section{Reaction times: numerical details}
\label{sec_ap:numerical}
\setcounter{equation}{0}

This section gives the numerical details corresponding to the simulation presented in section~\ref{sec:res_rt}.
This numerical example involves two equiprobable Gaussian categories, centered in $x^{\mu_1}=-3$ and $x^{\mu_1}=3$, with standard deviation $a^{\mu_1}=a^{\mu_2}=1.5$. The neuronal population (coding layer) is made of $N=10$ cells, with bell-shaped tuning curves,
\begin{equation}
f_i(x) = f_{\min} + (f_{\max}-f_{\min})\, \exp \left( - \frac{(x-x_i)^2}{2a_i^2}\right).
\label{S_eq:bell}
\end{equation}
The preferred stimuli $x_i$ of the neurons are initially equidistributed along the domain $[-6, 6]$. Before learning, each tuning curve has the same width ($a_i=2$). Minimal and maximal values of the firing rates are respectively set to $f_{\min}=0.001$ and $f_{\max}=5$.\\
During the learning phase, $100000$ stimuli are presented to the network, and both the weights $\mathbf{w}$ and 
the parameters of tuning curves (width and location) are optimized. The time window $\tau_a$ used during learning is equal to 1.\\
After learning, we look at the response of the network following the presentation of a stimulus, according to the diffusion model presented in Section~\ref{sec:rt}.
The simulation of this diffusion process is done as follows. We first generate a Poisson process by dividing the time interval  $[0, 3\tau_a]$ into 3000 bins. For a neuron $i$, each interval, of width $d\tau=\tau_a/1000$, receives a spike according to a Bernoulli law of parameter $f_i^{0}(x)\,d\tau$ ($d\tau$ being small, we thus get a Poisson process associated with each neuron). We then compute the temporal evolution of the output $\alpha_{\tau}$ as well as the time $\tau_d$ for which this quantity reaches one of the two bounds for the first time. In this numerical example, the bound $\gamma$ is set equal to 0.3. For each stimulus $x$, this process is run 10000 times, which makes it possible to have an estimate of the mean reaction time $\overline{\tau_d}(x)$. In the end, this operation is done for 20 stimuli equidistributed along a continuum ranging from $-4$ to $4$.

\clearpage
\pagebreak

\bibliographystyle{apalike_mod}
\bibliography{refs.bib}

\end{document}